\documentclass[prl,aps,twocolumn,preprintnumbers,amssymb,nobibnotes,nofootinbib,raggedbottom,10pt]{revtex4-1}
\usepackage{hyperref,graphicx,array,multirow,color,amsmath,mathrsfs,titlesec,shuffle,tikz}
\usetikzlibrary{calc,decorations,decorations.markings,decorations.pathmorphing,decorations.pathreplacing,shapes.geometric,patterns,shapes,decorations.pathmorphing,arrows}
\titleformat{\section}{\centering\normalsize\normalfont\bf}{\thesection}{0em}{}
\hypersetup{pdftitle={},pdfcreator={},linkcolor=[rgb]{0.15,0.35,0.75},colorlinks=true,citecolor=[rgb]{0.675,0,0.2},urlcolor=[rgb]{0.15,0.35,0.65}}
\newcommand{\eq}[1]{\vspace{-0.5pt}\begin{equation}#1\vspace{-0.5pt}\end{equation}}

\newcommand{\fwbox}[2]{\text{\makebox[#1][c]{$\hspace{-150pt}\displaystyle#2\hspace{-150pt}$}}}
\newcommand{\fwboxL}[2]{\text{\makebox[#1][l]{$#2$}}}
\newcommand{\fwboxR}[2]{\text{\makebox[#1][r]{$#2$}}}
\newcommand{\equivR}{\fwbox{14.5pt}{\hspace{-0pt}\fwboxR{0pt}{\raisebox{0.47pt}{\hspace{1.25pt}:\hspace{-4pt}}}=\fwboxL{0pt}{}}}
\newcommand{\equivL}{\fwbox{14.5pt}{\fwboxR{0pt}{}=\fwboxL{0pt}{\raisebox{0.47pt}{\hspace{-4pt}:\hspace{1.25pt}}}}}
\newcommand{\fig}[3]{\raisebox{#1}{\includegraphics[scale=#2]{#3}}}

\newcommand{\bigger}[1]{\raisebox{-0.95pt}{\scalebox{1.25}{$#1$}}}
\newcommand{\mi}{\raisebox{0.75pt}{\scalebox{0.75}{$\hspace{-0.5pt}\,-\,\hspace{-0.5pt}$}}}
\renewcommand{\pl}{\raisebox{0.75pt}{\scalebox{0.75}{$\hspace{-0.5pt}\,+\,\hspace{-0.5pt}$}}}
\renewcommand{\phi}{\varphi}

\renewcommand{\hat}{\widehat}
\renewcommand{\tilde}{\widetilde}
\newcommand{\ab}[1]{\langle #1\rangle}
\renewcommand{\sb}[1]{[ #1]}

\renewcommand{\r}[1]{{\color{hred}#1}}
\renewcommand{\b}[1]{{\color{hblue}#1}}
\newcommand{\g}[1]{{\color{hgreen}#1}}

\definecolor{nmhv}{rgb}{0.95,0.55,0.55}
\definecolor{mhvblue}{rgb}{0.6,0.6,0.7765}
\definecolor{ampgrey}{rgb}{0.9,0.9,0.9}
\definecolor{hblue}{rgb}{0,0,0.575}
\definecolor{hred}{rgb}{0.575,0.0,0.225}
\definecolor{hgreen}{rgb}{0.0,0.4,0.2}
\definecolor{hteal}{rgb}{0.0,0.545,0.7451}
\definecolor{perm}{rgb}{0.1,0.45,0.85}
\definecolor{unord}{rgb}{0,0,0}
\definecolor{ord}{rgb}{0,0,0.575}
\definecolor{anchorLeg}{rgb}{0.575,0.0,0.225}
\definecolor{fRed}{rgb}{0.48,0.02824,0.18824}

\begin{document}
\title{\texorpdfstring{Gauge-Invariant Double-Copies via Recursion\\[-22pt]~}{Gauge-Invariant Double-Copies via Recursion}}
\author{Jacob~L.~Bourjaily}%\email{bourjaily@psu.edu}
\author{Nikhil~Kalyanapuram}%\email{nkalyanapuram@psu.edu}
\author{Kokkimidis~Patatoukos}%\email{kzp326@psu.edu}
\author{Michael~Plesser}%\email{mkp5771@psu.edu}
\author{Yaqi~Zhang}%\email{yjz5289@psu.edu}
\affiliation{Institute for Gravitation and the Cosmos, Department of Physics,\\Pennsylvania State University, University Park, PA 16802, USA}
%%%%%%%%%%%%%%%%%%%%%%%%%%%%%%%%%%%%%%%%%%%%%%%%%%%%%%%%%%%%%%%%%%%%%%%%%%%%%%%%%%%%%%%%%%%

%%%%%%%%%%%%%%%%%%%%%%%%%%%%%%%%%%%%%%%%%%%%%%%%%%%%%%%%%%%%%%%%%%%%%%%%%%%%%%%%%%%%%%%%%%%
\begin{abstract} 
We prove that all tree-level amplitudes in pure (super-)gravity can be expressed as term-wise, gauge-invariant double-copies of those of pure (super-)Yang-Mills obtained via BCFW recursion. These representations are far from unique: varying the recursive scheme leads to a wide variety of distinct, but equally valid representations of gravitational amplitudes, all realized as double-copies.\\[-12pt]
\vspace{-10pt}
\end{abstract}
\maketitle
%%%%%%%%%%%%%%%%%%%%%%%%%%%%%%%%%%%%%%%%%%%%%%%%%%%%%%%%%%%%%%%%%%%%%%%%%%%%%%%%%%%%%%%%%%%

%%%%%%%%%%%%%%%%%%%%%%%%%%%%%%%%%%%%%%%%%%%%%%%%%%%%%%%%%%%%%%%%%%%%%%%%%%%%%%%%%%%%%%%%%%%
\vspace{-15pt}\section{Introduction}\label{introduction_section}\vspace{-14pt}
%%%%%%%%%%%%%%%%%%%%%%%%%%%%%%%%%%%%%%%%%%%%%%%%%%%%%%%%%%%%%%%%%%%%%%%%%%%%%%%%%%%%%%%%%%%
%

The rich connections between scattering amplitudes in gauge theory and gravity has been a source of tremendous progress in our understanding of both theories. Among the most seminal of these is so-called color-kinematic (or `BCJ') duality \cite{Bern:2008qj}, which states that gravitational scattering amplitudes may be represented as `double-copies' of those of Yang-Mills theory, provided the latter is represented in terms of color-kinematic satisfying (`dual') numerators with denominators (typically) built from scalar $\phi^3$ field theory (see e.g.~\mbox{\cite{Bern:2010ue,Bern:2019prr,Adamo:2022dcm,Bern:2023zkg}}). The existence of such numerators was first conjectured, but can be proven at tree-level in a number of ways \mbox{\cite{Bjerrum-Bohr:2010pnr,Mafra:2011kj,Bjerrum-Bohr:2016axv,Mizera:2016jhj,Mizera:2019blq}}, with much evidence suggesting that color-kinematic duality should continue to the level of loop integrands (see e.g.~\mbox{\cite{Carrasco:2011mn,Carrasco:2012ca,Bjerrum-Bohr:2013iza,Monteiro:2013rya,He:2015wgf,Mogull:2015adi,Johansson:2017bfl,Bern:2017yxu,Bern:2017ucb,Edison:2022jln,Porkert:2022efy}}). The potential form, structure, and scope of these numerators, as well as the theoretical origins of this story more generally have been the subject of a great deal of research (see e.g.~\mbox{\cite{Monteiro:2011pc,Bjerrum-Bohr:2012kaa,Chen:2019ywi,Borsten:2020zgj,Ahmadiniaz:2021ayd,Godazgar:2022gfw,Bonezzi:2022bse}}).

Prior to the discovery of color-kinematic duality, on-shell recursion relations \cite{Roiban:2004yf,BCF,BCFW} for tree-level scattering amplitudes led to similarly great leaps in our understanding of gravitational and gauge-theory amplitudes  \cite{Bedford:2005yy,Cachazo:2005ca,Benincasa:2007qj,Bjerrum-Bohr:2005xoa,Spradlin:2008bu,Elvang:2007sg,Drummond:2009ge,Mason:2009afn}. Some of this work connected directly to results derived from string and twistor string theory (e.g.~\cite{Berends:1988zp,Bern:1993wt,Bern:1999ji,Giombi:2004ix,Nair:2005iv}).

In this work, we show that BCFW recursion relations directly lead to representations of amplitudes in color-dressed Yang-Mills theory (YM) and gravity (GR) that may be expressed in the form
\vspace{-5pt}\eq{\begin{split}
\fwboxL{190pt}{\hspace{-15pt}\mathcal{A}^{\text{YM}}(1,\cdots,n)=\hspace{-5pt}\sum_{\b{\vec{a}}\in\mathfrak{S}\!(\!\b{A}\!)}\hspace{-2pt}\sum_{\Gamma}\fwbox{62.5pt}{\frac{c_{\r{\alpha\beta}}^{\hspace{2.2pt}\b{\vec{a}}}\hspace{3pt}n\big(\Gamma_{\hspace{-2pt}\r{\alpha\beta}}^{\hspace{0.5pt}\b{\vec{a}}}\big)}{D{\big(\Gamma_{\hspace{-2pt}\r{\alpha\beta}}^{\hspace{0.5pt}\b{\vec{a}}}\big)}}}\delta^{2\!\times\!2}\!\big(\lambda\!\cdot\!\tilde\lambda\big)}\\
\fwboxL{190pt}{\hspace{-15pt}\mathcal{A}^{\text{GR}}(1,\cdots,n)=\hspace{-5pt}\sum_{\b{\vec{a}}\in\mathfrak{S}\!(\!\b{A}\!)}\hspace{-2pt}\sum_{\Gamma}\fwbox{64.5pt}{\frac{{n\big(\Gamma_{\hspace{-2pt}\r{\alpha\beta}}^{\hspace{0.5pt}\b{\vec{a}}}\big)}\,{n\big(\Gamma_{\hspace{-2pt}\r{\alpha\beta}}^{\hspace{0.5pt}\b{\vec{a}}}\big)}}{D{\big(\Gamma_{\hspace{-2pt}\r{\alpha\beta}}^{\hspace{0.5pt}\b{\vec{a}}}\big)}}}\delta^{2\!\times\!2}\!\big(\lambda\!\cdot\!\tilde\lambda\big)}\\[-8pt]
\end{split}\label{yang_mills_and_gr_amps}\vspace{-4pt}}
for any choice $\{\r{\alpha},\r{\beta}\}\!\subset\![n]$ of the external legs, where $\b{A}\!\equivR\![n]\backslash\{\r{\alpha},\r{\beta}\}$.  For Yang-Mills theory, the form (\ref{yang_mills_and_gr_amps}) will be seen to be somewhat quixotic, as the color-kinematic dual `numerators' will be simply \emph{defined} to be the product of $D{\big(\Gamma_{\hspace{-2pt}\r{\alpha\beta}}^{\hspace{0.5pt}\b{\vec{a}}}\big)}$ and a more familiar gauge-invariant, on-shell function \cite{ArkaniHamed:book}; as such, the real novelty arises in the identification of the \emph{denominators} $D{\big(\Gamma_{\hspace{-2pt}\r{\alpha\beta}}^{\hspace{0.5pt}\b{\vec{a}}}\big)}$, which we define recursively. It is worth pointing out that because the color-factors appearing in (\ref{yang_mills_and_gr_amps}) are entirely independent, these numerators are only `color-kinematic dual' in a rather trivial sense: neither satisfies any identities.

The existence of formulae such as (\ref{yang_mills_and_gr_amps}) follows from the on-shell diagrammatic interpretation of BCFW recursion in YM. Ignoring factors of color and momentum conservation, the double-copy follows from the fact that for any primitive\footnote{A \emph{primitive} diagram is one involving only three-point amplitudes at its vertices. Any diagram with higher-point amplitudes can be expanded as a sum of primitives via BCFW recursion.} on-shell diagram $\Gamma$, the on-shell functions $f_\Gamma$ of gravity and Yang-Mills differ by a simple factor depending on the graph: 
\vspace{-3pt}\eq{\mathfrak{f}_\Gamma^{\text{GR}}=J(\Gamma)\big(\mathfrak{f}_\Gamma^{\text{YM}}\big)^2\,.\label{basic_relationship}\vspace{-3pt}}
This general fact (see e.g.~\cite{Heslop:2016plj,Herrmann:2016qea,Paranjape:2022ymg,Trnka:2020dxl,Brown:2022wqr}) is a simple consequence of the definition of an on-shell function and the relationship between the 3-particle $S$-matrices of the two theories: an on-shell function may be defined as the product of amplitudes evaluated on the \emph{residue} $1/J(\Gamma)$ of the scalar graph which puts all internal lines on-shell; squaring an on-shell function in YM gives the correct product of 3-particle amplitudes in GR, but squares also $1/J(\Gamma)$---which must be corrected by the numerator of (\ref{basic_relationship}).

%%%%%%%%%%%%%%%%%%%%%%%%%%%%%%%%%%%%%%%%%%%%%%%%%%%%%%%%%%%%%%%%%%%%%%%%%%%%%%%%%%%%%%%%%%%
\vspace{-15pt}\section{Tree-Level, On-Shell Recursion for YM and GR}\label{bcfw_for_ym_and_gr}\vspace{-14pt}
%%%%%%%%%%%%%%%%%%%%%%%%%%%%%%%%%%%%%%%%%%%%%%%%%%%%%%%%%%%%%%%%%%%%%%%%%%%%%%%%%%%%%%%%%%%
The starting point for on-shell (`BCFW') recursion \mbox{\cite{BCFW}} is to consider an amplitude as meromorphic function of external momenta and deform the momenta of any two particles labelled $\{\r{\alpha},\r{\beta}\}$ according to
\vspace{-4pt}\eq{\fwbox{0pt}{\hspace{-14pt}p_\r{\alpha}\!\mapsto\!\hat{p}_{\r\alpha}(\g{z})\!\equivR\!p_\r{\alpha}{+}\g{z}\,\lambda_{\r{\alpha}}\tilde\lambda_{\r{\beta}},\;\;
p_{\r\beta}\!\mapsto\!\hat{p}_{\r\beta}(\g{z})\!\equivR\!p_{\r\beta}{-}\g{z}\,\lambda_{\r{\alpha}}\tilde\lambda_{\r{\beta}},}\label{bcfw_shifts_defined}\vspace{-4pt}}
where $p_a\!\equivL\!\lambda_a\tilde\lambda_a$ are spinor-helicity variables \cite{vanderWaerden:1929}; this deformation preserves momentum conservation and keeps each particle on-shell. Note that the choice $\{\r{\alpha},\r{\beta}\}$ is distinct from the choice $\{\r{\beta},\r{\alpha}\}$: they differ by parity. At tree-level, amplitudes have poles at finite $\g{z}$ corresponding to factorization channels---the residues of which we may represent diagrammatically as
\eq{\fwbox{0pt}{\hspace{-10pt}\fig{-30pt}{1}{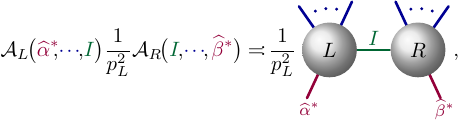}}\label{generic_factorization_diagram}}
where $1/p_L^2$ is the off-shell propagator being cut, with the left/right amplitudes evaluated with $p_{\r{\alpha^*}\!\!,\r{\beta^*}}\!\equivR\!p_{\r{\alpha},\r{\beta}}(\g{z^*})$ on the location of the pole $\g{z^*}\!\!=\!p_L^2/\langle\r{\alpha}|(\b{p_{L}})|\r{\beta}]$ and summed over the states $\g{I}$ that can be exchanged. Importantly, the deformed legs must necessarily be on opposite sides of the factorization channel for the simple reason that $\hat{p}_{\r\alpha}\!{+}\hat{p}_{\r\beta}=p_{\r\alpha}\!{+}p_{\r{\beta}}$ is $\g{z}$-independent. 

Provided there are no poles at infinity, Cauchy's theorem allows us to write an amplitude as a sum over all residues of the form (\ref{generic_factorization_diagram}) \cite{ArkaniHamed:2008yf}. For YM or GR, this will be the case provided the deformed momenta are chosen judiciously according to their helicity (see e.g.~\cite{Cohen:2010mi}), while amplitudes in maximally supersymmetric ($\mathcal{N}\!=\!4$) YM (`sYM') or ($\mathcal{N}\!=\!8$) GR (`sGR') will be free of poles at infinity regardless of which legs are chosen. Because tree-level amplitudes in pure (or any degree of less supersymmetric) YM/GR are identical to those of sYM/sGR for appropriately restricted sets of external states, we may therefore consider the case of maximally supersymmetric YM/GR without loss of generality \cite{ArkaniHamed:2008gz}, ensuring that amplitudes are free of poles at infinity for any choice of legs $\{\r{\alpha},\r{\beta}\}$.

Notice that the channels (\ref{generic_factorization_diagram}) allow for arbitrary distributions of the other $(n\mi2)$ legs $\b{A}\!\equivR\![n]\backslash\{\r{\alpha},\r{\beta}\}$. Thus, on-shell recursion results in a sum of terms of the form
\eq{\fwboxL{200pt}{\hspace{-15pt}\mathcal{A}=\hspace{-0pt}\fwbox{14pt}{\sum_{\substack{\b{\vec{a}}\in\mathfrak{S}(\b{A})\\\fwbox{0pt}{(\b{\vec{a}_L}\!,\b{\vec{a}_R)=\b{\vec{a}}}}}}}\mathcal{A}_L\!\big(\r{\hat{\alpha}^*}\!\!\!,\b{\vec{a}_L},\!\g{I}\big)\frac{1}{p_{\r{\alpha}\,\b{a_L}}^2\!\!\!}\mathcal{A}_R\!\big(\g{I}\!,\!\b{\vec{a}_R},\!\r{\hat{\beta}^*}\big)\!\!\!\equivL\fwbox{14pt}{\sum_{\b{\vec{a}}\in\mathfrak{S}(\b{A})}}\mathcal{A}\big(\r{\alpha},\b{\vec{a}},\r{\beta}\big)\!,}\label{partial_amplitude_seed_recursion}}
where $\mathcal{A}\big(\r{\alpha},\b{\vec{a}},\r{\beta}\big)\!\equivR\!\mathcal{A}\big(\r{\alpha},\b{a_1},\b{\cdots},\b{a_{\text{-}1}},\r\beta\big)$ are \emph{partial amplitudes} involving external momenta with \emph{specific ordering}.

As amplitudes in color-dressed YM and gravity are fully permutation-invariant (due to Bose symmetry), any choice of legs $\{\r{\alpha},\r{\beta}\}$ may be taken; and any particular ordering of the other legs $\b{\vec{a}}\!\in\!\mathfrak{S}(\b{A})$ will suffice to generate the full amplitude upon summation over permutations of the labels $\b{\vec{a}}$. Thus we may without loss of generality focus our attention on the determination of the partial amplitude $\mathcal{A}\big(\r{1},\b{2},\b{\cdots},\b{n\mi1},\r{n}\big)$.

It is important to note that in \emph{neither theory} {is the partial amplitude unique}: not only does it depend on the legs chosen, but also the specific sequence of choices made for iterated recursion. (This may seem surprising for YM, as it is common to consider `color-stripped' partial-amplitudes (`primitives'), which do enjoy many scheme-independent properties.)

One particularly convenient recursion scheme would be to always choose the first and last leg of every iteratively recursed amplitude, and use the same parity of bridge at each stage of recursion. In the case of Yang-Mills, this results in partial amplitudes dressed by the color-factors appearing in the familiar `DDM' representation \cite{DelDuca:1999rs}. Letting $\mathcal{A}^{\text{YM}}\big(\r{\alpha},\b{\vec{a}},\r{\beta}\big)\equivL c_{\r{\alpha\beta}}^{\hspace{2.2pt}\b{\vec{a}}}A^{\text{YM}}\big(\r{\alpha},\b{\vec{a}},\r{\beta}\big)$, it is easy to see that the recursion (\ref{partial_amplitude_seed_recursion}) separates color and kinematics cleanly so that---upon recursing iteratively down to factorizations involving only three-point amplitudes---we find
\vspace{-10pt}\eq{\begin{split}c_{\r{\alpha\beta}}^{\hspace{2.2pt}\b{\vec{a}}}&\equivR\hspace{-3pt}\sum_{e_i}c^{{\color{anchorLeg}\alpha},{\color{ord}a_1},e_1}c^{e_1,{\color{ord}a_2},e_2}\cdots c^{e_{\text{-}1},{\color{ord}a_{\text{-}1}},{\color{anchorLeg}\beta}} \text{}\\[-12pt]
\end{split}\label{ddm_color_factor_defined}}
where $c_{a\,b}^{\,\,\,c}$ are the structure constants of some Lie algebra (into which we may freely absorb any coupling constant), and $\{a,b,c\}$ are (adjoint) color-labels for the gluons. These color tensors are all independent under Jacobi-relations, and the so-called `{primitive}' ordered amplitudes of YM turn out to be gauge-invariant, local, dihedrally symmetric, and to enjoy KK relations. (All of these properties can be deduced from the Jacobi identity and Bose symmetry of color-dressed amplitudes alone.) Besides gauge-invariance, none of these properties will be enjoyed by the partial amplitudes of gravity---the meaning of which will depend strongly on how recursion is implemented (analogously to color tensors for YM). 

Because this recursion scheme results in the same color-factor $c_{\r{\alpha\beta}}^{\hspace{2.2pt}\b{\vec{a}}}$ coefficient for every term, it is common to factor it out entirely and focus on the \emph{color-stripped} partial amplitude primitive $A^{\text{YM}}\big(\r{1},\b{2},\b{\cdots},\b{n\mi1},\r{n}\big)$.

\begin{table*}\vspace{-8pt}\caption{On-shell, gauge-invariant contributions to the 6-point NMHV partial amplitudes of Yang-Mills and gravity.\\ 
The Grassmann $\delta$-functions $\delta^{3\!\times\!4}\!\big(C_i\!\!\cdot\!\tilde\eta\big)$ appearing in these numerators are defined in (\ref{nmhv_deltas_defined}).\label{six_point_nmhv_factors}}\vspace{-15pt}$$\fig{-50pt}{1}{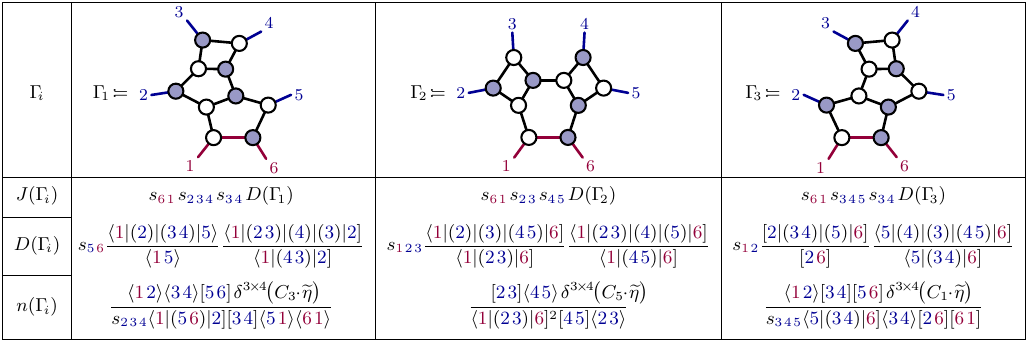}\vspace{-24pt}$$\end{table*}

%%%%%%%%%%%%%%%%%%%%%%%%%%%%%%%%%%%%%%%%%%%%%%%%%%%%%%%%%%%%%%%%%%%%%%%%%%%%%%%%%%%%%%%%%%%
\vspace{-15pt}\section{On-Shell Diagrammatics of YM and GR}\label{on_shell_diagrams_in_gr_ym}\vspace{-14pt}
%%%%%%%%%%%%%%%%%%%%%%%%%%%%%%%%%%%%%%%%%%%%%%%%%%%%%%%%%%%%%%%%%%%%%%%%%%%%%%%%%%%%%%%%%%%
For color-stripped partial amplitudes in YM, there exists a powerful diagrammatic manifestation of recursion relations following from the simple fact that 
\vspace{-8pt}\eq{\fwbox{0pt}{\hspace{-10pt}\fig{-32pt}{1}{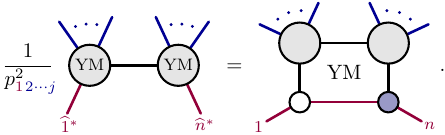}}\vspace{-6pt}\label{factorization_equals_box_for_ym}}
Here, we have introduced `flat' vertices to denote \emph{ordered} partial amplitude primitives. The right hand side represents an \emph{on-shell function} of YM: the product of (color-stripped) amplitudes at the vertices, summing over all the on-shell, internal states that can be exchanged between them. These functions are extremely well understood: they are classified combinatorially, and all their functional relations can be understood to arise homologically from an auxiliary Grassmannian `positive' geometry (see e.g.~\cite{ArkaniHamed:book,Bourjaily:2012gy}).

Applying recursion successively results in a representation of color-stripped YM partial amplitudes as sums over on-shell functions encoded by \emph{specific} on-shell diagrams $\{\Gamma\}$
\vspace{-4pt}\eq{\hspace{-10pt}{A}^{\text{YM}}\big(\r{1},\b{2},\b{\cdots},\b{n\mi1},\r{n}\big)=\sum_\Gamma \mathfrak{f}^{\text{YM}}_{\Gamma}\!\equivL\,\,\sum_{\Gamma}\,\,\hat{\mathfrak{f}}^{\text{YM}}_\Gamma\,\,\delta^{2\!\times\!2}\!\big(\lambda\!\cdot\!\tilde\lambda\big)\hspace{-10pt}\vspace{-5pt}}
where the sum is over on-shell diagrams $\{\Gamma\}$ of the form (\ref{factorization_equals_box_for_ym}) involving \emph{exclusively} three-point vertices. The N${}^k$MHV-degree of an amplitude is encoded by the graph according to $k{=}2\,n_B{+}n_W\,{-}n_I{-}2$, where $n_B(n_W)$ denotes the number of blue(white) vertices and $n_I$ the number of internal lines. Even for MHV amplitudes, there are vastly more on-shell \emph{diagrams} than on-shell functions---as diagrams related by mergers and square moves leave on-shell functions unchanged \emph{in YM} \cite{ArkaniHamed:book}. On-shell diagrams in gravity enjoy only the square move as an (un-modified) equivalence relation \cite{Heslop:2016plj}.

%%%%%%%%%%%%%%%%%%%%%%%%%%%%%%%%%%%%%%%%%%%%%%%%%%%%%%%%%%%%%%%%%%%%%%%%%%%%%%%%%%%%%%%%\%%%
%\newpage
\vspace{-15pt}\section{Color-Kinematic \emph{Denominators} for Gravity}\label{bcj_for_gravity}\vspace{-14pt}
%%%%%%%%%%%%%%%%%%%%%%%%%%%%%%%%%%%%%%%%%%%%%%%%%%%%%%%%%%%%%%%%%%%%%%%%%%%%%%%%%%%%%%%%%%%

For amplitudes in GR, there is no simple analogue of (\ref{factorization_equals_box_for_ym}) (see e.g.~\cite{Heslop:2016plj,Herrmann:2016qea}); but, supposing that there is some diagrammatic representation for partial amplitudes in GR in terms of on-shell diagrams of YM, we may recursively conclude that
\vspace{-8pt}\eq{\fwbox{0pt}{\hspace{-10pt}\fig{-24pt}{1}{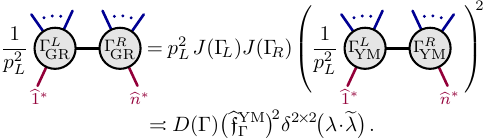}}\vspace{-4pt}\label{factorization_vs_box_for_gr}}
The fact that BCFW recursion for GR can be expressed in the form (\ref{factorization_vs_box_for_gr}) is reasonably well known \mbox{\cite{Heslop:2016plj,Herrmann:2016qea,Trnka:2020dxl,Paranjape:2022ymg,Brown:2022wqr}}. The precise form of $D(\Gamma)$ depends both on the graph $\Gamma$ and the recursion scheme followed. If we always choose the first and last labels for subsequent recursion, so that every diagram appearing is of the form $\Gamma_{\!\!\!\r{\alpha}\,\r{\beta}}^{\,\b{\vec{a}}}\!=\!\Gamma_{\!\!\!L}\big[\r{\hat{\alpha}^*}\!\!\!,\b{\vec{a}_L},\r{I}\big]\bigger{\otimes}\Gamma_{\!\!\!R}\big[\r{\r{I},\b{\vec{a}_R},\hat{\beta}^*}\big]$, then $D(\Gamma)$ will be
\vspace{-6pt}\eq{D\big(\Gamma_{\r{\alpha}\,\r{\beta}}^{\,\,\,\b{\vec{a}}}\big)=p^2_{\r{\alpha\,\b{a_L}}}D\big(\Gamma_{\!\!\!\r{\hat{\alpha}^*}\,\r{I}}^{\,\b{\vec{a}_L}}\big)D\big(\Gamma_{\substack{\\[-2pt]\!\!\!\r{I}\,\,\,\,\r{\hat{\beta}^*}\!\!}}^{\b{\vec{a}_R}}\big)\,.\label{default_scheme_denominators}\vspace{-6pt}}

We call these factors color-kinematic \emph{denominators} because if we let $n(\Gamma)\!\equivR\!D(\Gamma)\,\,\hat{\mathfrak{f}}_\Gamma^{\text{YM}}$ then individual terms appearing in the recursion of an amplitude in YM take the form 
\vspace{-12pt}\eq{\fwbox{140pt}{\fwbox{50.5pt}{\frac{c_{\r{\alpha\beta}}^{\hspace{2.2pt}\b{\vec{a}}}\hspace{3pt}n\big(\Gamma_{\hspace{-2pt}\r{\alpha\beta}}^{\hspace{0.5pt}\b{\vec{a}}}\big)}{D{\big(\Gamma_{\hspace{-2pt}\r{\alpha\beta}}^{\hspace{0.5pt}\b{\vec{a}}}\big)}}}=c_{\r{\alpha\beta}}^{\hspace{2.2pt}\b{\vec{a}}}\hspace{0pt}\,\,\hat{\mathfrak{f}}^{\,\text{YM}}_{\Gamma_{\hspace{-2pt}\r{\alpha\beta}}^{\hspace{0.5pt}\b{\vec{a}}}}}\vspace{-8pt}}
while terms in gravity are given by the \emph{double-copy}
\vspace{-5pt}\eq{\fwbox{140pt}{\hspace{-0pt}\fwbox{65.5pt}{\frac{n\big(\Gamma_{\hspace{-2pt}\r{\alpha\beta}}^{\hspace{0.5pt}\b{\vec{a}}}\big)\hspace{3pt}n\big(\Gamma_{\hspace{-2pt}\r{\alpha\beta}}^{\hspace{0.5pt}\b{\vec{a}}}\big)}{D{\big(\Gamma_{\hspace{-2pt}\r{\alpha\beta}}^{\hspace{0.5pt}\b{\vec{a}}}\big)}}}\!=\!D\big(\Gamma_{\hspace{-2pt}\r{\alpha\beta}}^{\hspace{0.5pt}\b{\vec{a}}}\big)\hspace{-2pt}\,\,\big(\,\hat{\mathfrak{f}}^{\,\text{YM}}_{\Gamma_{\hspace{-2pt}\r{\alpha\beta}}^{\hspace{0.5pt}\b{\vec{a}}}}\big)^{\!\!2}\,.}\vspace{-8pt}}
%

%%%%%%%%%%%%%%%%%%%%%%%%%%%%%%%%%%%%%%%%%%%%%%%%%%%%%%%%%%%%%%%%%%%%%%%%%%%%%%%%%%%%%%%%%%%
\vspace{-20pt}\section{Illustrations of On-Shell Double-Copies}\label{exempli_gratia}\vspace{-14pt}
%%%%%%%%%%%%%%%%%%%%%%%%%%%%%%%%%%%%%%%%%%%%%%%%%%%%%%%%%%%%%%%%%%%%%%%%%%%%%%%%%%%%%%%%%%%
Arguably the simplest amplitudes in either theory are the so-called (N${}^{k=0}$)MHV amplitudes \cite{Parke:1986gb,Berends:1988zp,Nair:2005iv}. On-shell recursion results in a single term for ordered partial amplitudes in either theory:
\vspace{-12pt}\eq{\begin{split}\hspace{-20pt}\mathcal{A}^{\text{YM}}_{\text{MHV}}(\r{1},\b{\cdots},\r{n})&=c_{\r{1}\,\,\r{n}}^{\,\,\,\b{\vec{a}}}\frac{\delta^{2\!\times\!4}\big(\lambda\!\cdot\!\tilde\eta\big)\,\,\fwboxL{20pt}{\delta^{2\!\times\!2}\!\big(\lambda\!\!\cdot\!\!\tilde\lambda\big)}}{\ab{\r{1}\,\b{2}}\ab{\b{2\,3}}\cdots\ab{\b{n\mi1}\,\r{n}}\ab{\r{n\,1}}}\\[-4pt]
&\fwboxL{0pt}{\hspace{-70pt}\equivL\! c_{\r{1}\,\,\r{n}}^{\,\,\,\b{\vec{a}}}\frac{n\big(\Gamma^{\text{MHV}}_{\!\!\!\r{1\,\,\,n}}\big)}{D\big(\Gamma^{\text{MHV}}_{\!\!\!\r{1\,\,\,n}}\big)}\,\delta^{2\!\times\!2}\!\big(\lambda\!\!\cdot\!\!\tilde\lambda\big)\,\fwboxL{0pt}{\!\!\equivL\!\!c_{\r{1}\,\,\r{n}}^{\,\,\,\b{\vec{a}}}\text{PT}(\r{1}\b{\cdots}\,\r{n})\,\delta^{2\!\times\!2}\!\big(\lambda\!\!\cdot\!\!\tilde\lambda),}}
\end{split}}
where the denominators are determined recursively. In the default recursion scheme (\ref{default_scheme_denominators}), we find\\[-14pt]
\eq{\begin{split}
\hspace{-20pt}D\big(\Gamma^{\text{MHV}}_{\!\!\!\r{1\,\,\,n}}\big)&\equivR p_{\b{n{-}1}\,\r{n}}^2\prod_{j=\b{4}}^{\b{n{-}1}}\frac{\langle\r{1}|(\b{2\cdots j\mi2})|(\b{j\mi1})|\b{j}\rangle}{\ab{\r{1}\,\b{j}}}\,,\hspace{-20pt}\\
\hspace{-20pt}n\big(\Gamma^{\text{MHV}}_{\!\!\!\r{1\,\,\,n}}\big)&\equivR D\big(\Gamma^{\text{MHV}}_{\!\!\!\r{1\,\,\,n}}\big)\frac{\delta^{2\!\times\!4}\big(\lambda\!\cdot\!\tilde\eta\big)}{\ab{\r{1}\,\b{2}}\ab{\b{2\,3}}\cdots\ab{\b{n\mi1}\,\r{n}}\ab{\r{n\,1}}}\,\hspace{-20pt}\\
\hspace{-20pt}&=\frac{\sb{\b{3\,2}}}{\ab{\b{2\,3}}\ab{\r{n\,1}}^2}\prod_{j=\b{4}}^{\b{n{-}1}}\frac{\langle\r{1}|(\b{2}\b{\cdots}\b{j\mi1})|\b{j}]}{\ab{\r{1}\,\b{j}}}\,\,\delta^{2\!\times\!4}\!\big(\lambda\!\cdot\!\tilde\eta\big);\hspace{-20pt}\\[-8pt]\end{split}}
this representation immediately allows us to write the corresponding expressions for GR as a double-copy:
\eq{\begin{split}\hspace{-20pt}\mathcal{A}^{\text{GR}}_{\text{MHV}}\big(\r{1},\b{\cdots},\r{n}\big)&=\frac{n\big(\Gamma^{\text{MHV}}_{\!\!\!\r{1\,\,\,n}}\big)\,n\big(\Gamma^{\text{MHV}}_{\!\!\!\r{1\,\,\,n}}\big)}{D\big(\Gamma^{\text{MHV}}_{\!\!\!\r{1\,\,\,n}}\big)}\delta^{2\!\times\!2}\!\big(\lambda\!\cdot\!\tilde\lambda\big)\hspace{-18pt}\\
\hspace{-20pt}&=D\big(\Gamma^{\text{MHV}}_{\!\!\!\r{1\,\,\,n}}\big)\big(\text{PT}(\r{1}\b{\cdots}\r{n})\big)^{\!\!2}\delta^{2\!\times\!2}\!\big(\lambda\!\cdot\!\tilde\lambda\big).\hspace{-18pt}
\end{split}\label{gr_mhv_default}}
Unlike the case of YM, these partial amplitudes in GR are non-cyclic and involve non-local poles. Only upon summing over all $(n\mi2)!$ orderings of $\{\b{2},\b{\cdots},\b{n\mi1}\}$ do we recover a local, permutation-invariant amplitude. We have checked that this formula agrees with the closed-form expression of Hodges \cite{Hodges:2011wm} through $n\!=\!12$ particles.

It is worth noting that this representation of MHV amplitudes in gravity (\ref{gr_mhv_default}) is \emph{identical} (upon a rotation of labels) to that found in \cite{Elvang:2007sg}. And as with \cite{Elvang:2007sg}, the use of the `bonus relations' stemming from the good large-$\g{z}$ behavior of amplitudes in GR \cite{ArkaniHamed:2008gz} allows us to re-write (\ref{gr_mhv_default}) as a sum over $(n\mi3)!$ terms \cite{Spradlin:2008bu}:
\vspace{-10pt}\eq{\begin{split}\\[-10pt]\hspace{-0pt}\hat{\mathcal{A}}^{\text{GR}}_{\text{MHV}}&=\sum_{\fwbox{40pt}{\b{\vec{a}}\!\in\!\!\mathfrak{S}\!(\b{3\hspace{-0.5pt},\!\!\cdots\!\!,\!n\text{-}1})}}\frac{\ab{\r{n}\,\r{1}}\ab{\g{2}\,\b{3}}}{\ab{\r{n}\,\g{2}}\ab{\r{1}\,\b{3}}}D\big(\!\Gamma^{\g{2}\b{\vec{a}}}_{\!\!\!\r{1}\,\r{n}}\big)\big(\text{PT}(\r{1}\,\g{2}\,\b{\vec{a}}\,\r{n})\big)^{\!\!2}\,.\end{split}\vspace{-15pt}\label{bonus_relation_formula}}

\setcounter{table}{2}\begin{table*}\vspace{-8pt}\caption{Alternative recursion schemata resulting in distinct ordered, partial amplitudes `$\mathcal{A}^{\text{GR}}\!\big(\r{1},\b{2},\b{3},\b{4},\b{5},\r{6}\big)$'.\label{six_point_mhv_variations}}\vspace{-18pt}$$\fig{-50pt}{1}{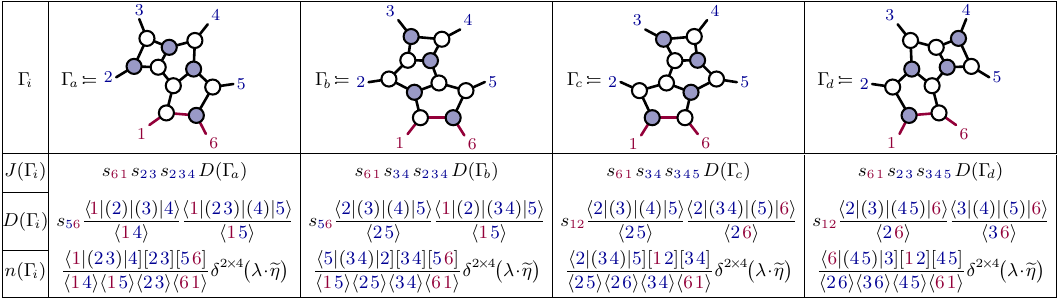}\vspace{-20pt}$$\end{table*}

For higher N$^{k\!>\!0}$MHV-degrees, on-shell recursion typically involves a sum over terms, each represented in YM by a particular, primitive on-shell diagram. The simplest non-trivial example is the 6-particle NMHV amplitude, which involves 3 terms to represent the ordered amplitude. Following the recursion scheme described above, the three on-shell diagrams $\{\Gamma_{\!\!1},\Gamma_{\!\!2}\,\Gamma_{\!\!3}\}$ that result are given in \mbox{Table \ref{six_point_nmhv_factors}}, where we have also indicated the numerators $n(\Gamma_{\!\!i})$ and denominators $D(\Gamma_{\!\!i})$ of each. Thus, we may write the ordered, partial NMHV amplitude primitive in YM as 
\vspace{-6pt}\eq{\fwboxL{200pt}{\hspace{-14pt}\mathcal{A}_{6,1}^{\text{YM}}\big(\r{1},\!\!\b{\cdots}\!,\r{n}\big)\!\!=\!c_{\r{1}\,\r{6}}^{\,\,\b{\vec{a}}}\!\!\left(\!\!\!\frac{n\big(\Gamma_{\!\!1}\big)}{D\big(\Gamma_{\!\!1}\big)}{+}\frac{n\big(\Gamma_{\!\!2}\big)}{D\big(\Gamma_{\!\!2}\big)}{+}\frac{n\big(\Gamma_{\!\!3}\big)}{D\big(\Gamma_{\!\!3}\big)}\!\!\right)\!\!\delta^{2\!\times\!2}\!\big(\!\lambda\!\!\cdot\!\!\tilde\lambda\big)}\vspace{-6pt}\label{6nmhv_ym}}
and the corresponding partial amplitude in GR as the double-copy
\vspace{-6pt}\eq{\fwboxL{200pt}{\hspace{-14pt}\mathcal{A}_{6,1}^{\text{GR}}\big(\r{1},\!\!\b{\cdots}\!,\r{n}\big)\!=\!\!\left(\!\!\!\frac{n\big(\Gamma_{\!\!1}\big)^{\!2}}{D\big(\Gamma_{\!\!1}\big)}{+}\frac{n\big(\Gamma_{\!\!2}\big)^{\!2}}{D\big(\Gamma_{\!\!2}\big)}{+}\frac{n\big(\Gamma_{\!\!3}\big)^{\!2}}{D\big(\Gamma_{\!\!3}\big)}\!\!\right)\!\!\delta^{2\!\times\!2}\!\big(\!\lambda\!\!\cdot\!\!\tilde\lambda\big)\!.}\vspace{-6pt}}
In both cases, these partial amplitudes must be summed over the $(n\mi2)!$ orderings of the legs $\{\b{2},\b{\cdots},\b{5}\}$. We have checked this expression against KLT \cite{Kawai:1985xq}.

The numerators listed in \mbox{Table \ref{six_point_nmhv_factors}} involve Grassmann $\delta$-functions involving the $\tilde\eta$'s which label the external states of each supermultiplet \cite{ArkaniHamed:2008gz}; they are defined by 
\vspace{-6pt}\eq{\begin{split}&\hspace{-15pt}\delta^{3\!\times\!4}\!\big(C_{a}\!\!\cdot\!\tilde\eta\big)\equivR\\
&\fwboxL{200pt}{\hspace{-15pt}\delta^{2\!\times\!4}\!\big(\lambda\!\cdot\!\tilde\eta\big)\delta^{1\!\times\!4}\!\big([a\,a\pl1]\tilde\eta_{a\,\text{{-}}1}\text{+}[a\pl1\,\,a\mi1]\tilde\eta_{a}{+}[a\,\mi1\,a]\tilde\eta_{a\text{+}1}\!\big)\,.}\\[-8pt]\end{split}\label{nmhv_deltas_defined}\vspace{-6pt}}
Notice that $\delta^{3\!\times\!4}\big(C_a\!\!\cdot\!\tilde\eta\big)$ is invariant under permutations of both the set $\{a\mi1,a,a\pl1\}$ and its complement. This will turn out to have important consequences as we discuss in the forthcoming work \cite{structureOfOnShellGR}.  

Although the expression (\ref{6nmhv_ym}) may seem unusual, it is worth observing that, for example,
\vspace{-4pt}\eq{\frac{n\big(\Gamma_{\!\!2}\big)}{D\big(\Gamma_{\!\!2}\big)}=\frac{\delta^{2\!\times\!4}\!\big(\lambda\!\cdot\!\tilde\eta\big)\delta^{1\!\times\!4}\!\big(\sb{\b{5}\r{6}}\tilde\eta_{\b{4}}\pl\sb{\r{6}\b{4}}\tilde\eta_{\b{5}}\pl\sb{\b{45}}\tilde\eta_\r{6}\big)}{\sb{\b{5}\r{6}}\ab{\r{1}\b{2}}\langle\b{3}|(\b{45})|\r{6}]s_{\r{1}\b{23}}[\b{4}|(\b{5}\r{6})\r{1}\rangle\ab{\b{23}}\sb{\b{45}}},\vspace{-5pt}}
is simply the (momentum-space version of the) familiar $R$-invariant $\mathcal{R}\big[\r{1},\b{3},\b{4},\b{5},\r{6}\big]$ (see e.g.~\cite{Drummond:2008vq}). 

Up to minor conventional differences, the recursion scheme used to construct denominators (\ref{default_scheme_denominators}) generally reproduces the form of amplitudes in GR as they were derived in \cite{Drummond:2009ge}. We have implemented in this in \textsc{Mathematica} and have verified agreement against KLT \cite{Kawai:1985xq} through the 10-particle N${}^3$MHV amplitude. These tools will be made available in a forthcoming, public package for tree amplitudes more generally \cite{toolsForTrees} (see also~\cite{Bourjaily:2010wh}).

\setcounter{table}{1}\begin{table}[t]\vspace{-8pt}\caption{Alternative recursion schemata resulting in distinct ordered, partial amplitudes `$\mathcal{A}^{\text{GR}}\!\big(\r{1},\b{2},\b{3},\b{4},\r{5}\big)$'.\label{five_point_mhv_variants}}\vspace{-16pt}$$\fig{-50pt}{1}{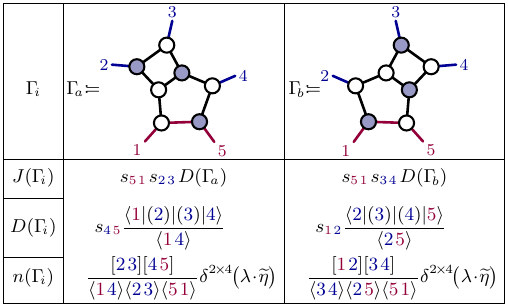}\vspace{-24pt}$$\end{table}\setcounter{table}{3}

%\newpage
%%%%%%%%%%%%%%%%%%%%%%%%%%%%%%%%%%%%%%%%%%%%%%%%%%%%%%%%%%%%%%%%%%%%%%%%%%%%%%%%%%%%%%%%%%%
\vspace{-15pt}\section{Non-Uniqueness of Dual Denominators}\label{nonuniqueness}\vspace{-14pt}
%%%%%%%%%%%%%%%%%%%%%%%%%%%%%%%%%%%%%%%%%%%%%%%%%%%%%%%%%%%%%%%%%%%%%%%%%%%%%%%%%%%%%%%%%%%
%
As emphasized above, even restricting ourselves to successively choosing the first last legs for all iterated recursions, variability emerges from the chiral asymmetry of the BCFW deformation (\ref{bcfw_shifts_defined}). Even for 5 particles, choosing $\{\r{\alpha},\r{\beta}\}$ to be $\{\r{1},\r{5}\}$ versus $\{\r{5},\r{1}\}$ (conjugating the shifting rule) results in two distinct diagrams $\Gamma_{\!\!a}$ and $\Gamma_{\!\!b}$ and correspondingly \emph{distinct} primitives $n(\Gamma_{\!\!i})^2/D(\Gamma_{\!\!i})$
\vspace{-6pt}\eq{\begin{split}\mathcal{A}_a^{\text{GR}}\big(\r{1},\b{\cdots},\r{5}\big)
&\equivR% \frac{n\big(\Gamma_{\!\!a}\big)^{\!2}}{D\big(\Gamma_{\!\!a}\big)}\delta^{2\!\times\!2}\!\big(\lambda\!\cdot\!\tilde\lambda\big)
\frac{\ab{\r{1}|(\b{2})|(\b{3})|\b{4}}\sb{\b{4}\,\r{5}}\fwboxL{26pt}{\,\delta^{2\!\times\!8}\!\big(\lambda\!\cdot\!\tilde\eta\big)}}{\ab{\r{1}\,\b{4}}\ab{\b{4}\,\r{5}}(\!\ab{\r{1}\,\b{2}}\ab{\b{2\,3}}\ab{\b{3\,4}}\ab{\r{5\,1}}\!)^{2}}\,\delta^{2\!\times\!2}\!\big(\lambda\!\cdot\!\tilde\lambda\big);\\
\mathcal{A}_b^{\text{GR}}\big(\r{1},\b{\cdots},\r{5}\big)
&\equivR% \frac{n\big(\Gamma_{\!\!b}\big)^{\!2}}{D\big(\Gamma_{\!\!b}\big)}\delta^{2\!\times\!2}\!\big(\lambda\!\cdot\!\tilde\lambda\big)=
\frac{\ab{\b{2}|(\b{3})|(\b{4})|\r{5}}\sb{\r{1}\,\b{2}}\fwboxL{26pt}{\,\delta^{2\!\times\!8}\!\big(\lambda\!\cdot\!\tilde\eta\big)}}{\ab{\r{1}\,\b{2}}\ab{\b{2}\,\r{5}}(\!\ab{\b{2\,3}}\ab{\b{3\,4}}\ab{\b{4\,5}}\ab{\r{5\,1}}\!)^{2}}\,\delta^{2\!\times\!2}\!\big(\lambda\!\cdot\!\tilde\lambda\big).\\[-20pt]
\end{split}\nonumber}
Nevertheless, it is easy to verify that the sum over their permuted images agree:
\eq{\mathcal{A}^{\text{GR}}\!=\!\!\!\!\fwbox{20pt}{\sum_{\fwboxL{20pt}{\b{\vec{a}}\!\in\!\mathfrak{S}([\b{2},\!\cdots\!,\b{4}])}}}\hspace{-2pt}\mathcal{A}^{\text{GR}}_a\big(\r{1},\b{a_1},\!\cdots\!,\b{a_{\text{-}1}},\r{5}\big)=\!\!\!\fwbox{20pt}{\sum_{\fwboxL{20pt}{\b{\vec{a}}\!\in\!\mathfrak{S}([\b{2},\!\cdots\!,\b{4}])}}}\hspace{-2pt}\mathcal{A}^{\text{GR}}_b\big(\r{1},\b{a_1},\!\cdots\!,\b{a_{\text{-}1}},\r{5}\big).\nonumber}

This variability only proliferates for higher multiplicity, as evidenced by the four examples for 6-point MHV given in \mbox{Table \ref{six_point_mhv_variations}}. More generally, the equivalence of expressions upon distinct variations gives powerful identities among not merely the partial amplitudes in YM, but even among individual on-shell functions appearing in N${}^{k\!>\!0}$MHV amplitudes. We will explore the scope of these possibilities and their consequences in a forthcoming work \cite{structureOfOnShellGR}.

%%%%%%%%%%%%%%%%%%%%%%%%%%%%%%%%%%%%%%%%%%%%%%%%%%%%%%%%%%%%%%%%%%%%%%%%%%%%%%%%%%%%%%%%%%%
\vspace{-12pt}\section{Conclusions and Future Directions}\vspace{-15pt}
%%%%%%%%%%%%%%%%%%%%%%%%%%%%%%%%%%%%%%%%%%%%%%%%%%%%%%%%%%%%%%%%%%%%%%%%%%%%%%%%%%%%%%%%%%%
The number of terms generated by BCFW to represent the $n$-particle N${}^k$MHV amplitude for a specific ordering is given by a Narayana number $\frac{1}{n\mi3}\binom{n\mi3}{k\pl1}\binom{n\mi3}{k}$. It is natural to suppose that including a sum over $(n\mi2)!$ permutations of leg labels would result in as many more terms in the expression for gravity. This turns out to not be the case---as evidenced, for example, by the more compact expression for MHV amplitudes (\ref{bonus_relation_formula}). 

Interestingly, for higher multiplicity and N${}^k$MHV-degree, considerations of Grassmannian geometry of the $\tilde\eta$-coefficients expose \emph{even more} symmetry than would result from mere permutation-invariance of amplitudes in GR \cite{structureOfOnShellGR}. For the 10-particle N${}^3$MHV amplitude, for example, we require only $343,\!252$ distinct superfunctions---about 20 times fewer than the na\"ive estimate of $(175\!\times\!8!)$. Moreover, the contributions that appear are found to satisfy a number of novel functional relations---some of which can be demonstrated using bonus relations or by equating formulae resulting from different recursion schemata, but we have also stumbled into yet further relations that remain to be understood. We explore some of these aspects of gravitational amplitudes in \cite{structureOfOnShellGR}. This additional, geometric structure hints at the possibility of a broader geometric story, perhaps analogous to the `gravituhedron' described in \cite{Trnka:2020dxl}.

While the existence of color-kinematic dual numerators remains conjectural beyond tree-level, there is a great deal of evidence from specific examples that the double-copy should generalize to loop-integrands (in some form or other) \mbox{\cite{Bern:2010yg,HenryTye:2010tcy,Carrasco:2011mn,Carrasco:2012ca,Bern:2012uf,Bjerrum-Bohr:2013iza,Monteiro:2013rya}}. In ref.~\cite{Heslop:2016plj}, Heslop and Lipstein gave evidence at one loop that the obvious extension of on-shell recursion for loop-integrands for sYM \cite{ArkaniHamed:2010kv} works also for sGR. It is natural to wonder if this works more generally, and if loop amplitude integrands for sGR continue to be generated as a double-copy of those for color-dressed sYM.

%%%%%%%%%%%%%%%%%%%%%%%%%%%%%%%%%%%%%%%%%%%%%%%%%%%%%%%%%%%%%%%%%%%%%%%%%%%%%%%%%%%%%%%%%%%
\vspace{-12pt}\section{Acknowledgments}\vspace{-15pt}
%%%%%%%%%%%%%%%%%%%%%%%%%%%%%%%%%%%%%%%%%%%%%%%%%%%%%%%%%%%%%%%%%%%%%%%%%%%%%%%%%%%%%%%%%%%
The authors gratefully acknowledge fruitful conversations with Enrico Herrmann and Jaroslav Trnka, and helpful comments and encouragement from JJ Carrasco, Henrik Johansson, Marcus Spradlin, and Radu Roiban. This project benefited from the hospitality of NORDITA during the workshop ``Amplifying Gravity at All Scales'', and was supported by an ERC Starting Grant \mbox{(No.\ 757978)}, a grant from the Villum Fonden \mbox{(No.\ 15369)}, and the US Department of Energy under contract DE-SC00019066.

\vspace{-14pt}
\providecommand{\href}[2]{#2}\begingroup\raggedright\endgroup

\end{document}